# Extension of Technology Acceptance Model by using System Usability Scale to assess behavioral intention to use e-learning

Anastasia Revythi and Nikolaos Tselios
ICT in Education Group, Dept. of Educational Sciences and Early Childhood Education
University of Patras
Rio, Patras, 26500, Greece
arevythi@gmail.com, nitse@ece.upatras.gr

**Abstract**

This study examines the acceptance of technology and behavioral intention to use learning management systems (LMS). In specific, the aim of this research is to examine whether students ultimately accept and use educational learning systems such as e-class and the impact of behavioral intention on their decision to use them. An extended version of technology acceptance model has been proposed and used by employing the System Usability Scale to measure perceived ease of use. 345 university students participated in the study and the data analysis was based on partial least squares method. The results were confirmed in most of the research hypotheses. In particular, social norm, system access and self-efficacy significantly affect behavioral intention to use. As a result, it is suggested that e-learning developers and stakeholders should focus on these factors to increase acceptance and effectiveness of learning management systems.

## Introduction

During the recent years, the development of information systems has been performed at a rapid pace. This ascertainment raises significant questions related to the acceptance of those systems. By Acceptance of Technology defined as the willingness of a user to use the technology and tools which have been developed to support it (Teo, 2011) is expressed. A significant body of research unveils that users' intention to use a system is affected primarily by their perceived usefulness and ease of use of it (Dasgupta, Granger, & Mcgarry, 2002).

Several researchers examined factors which influence people to accept and use an information system (Hsu & Chang, 2013; King & He, 2006). For these reasons, researchers of information systems at times develop and elaborate different techniques to be able to understand these factors and to predict the success of the systems, thus consequently improving their design.

## Conclusions

In this paper, the assessment of acceptance and behavioral intention to use LMS using a modified version of TAM was examined. It was found that the factor of self-efficiency appeared to have a significant impact not only on perceived ease of use, but also on behavioral intention. Furthermore, it should be stressed that this finding was put forward by Park (2009) who came to the conclusion that this factor affects behavioral intention the most, an effect which is also supported by the original TAM theory (Davis & Venkatesh, 2000). One possible explanation for the influence of self-efficacy could be interpreted through incentive theory as postulated in the work of Bandura (1994) and the theory of intrinsic motivation, which support that higher self-efficacy leads to better results in the learning process and in this case, the use of the e - class. Therefore, self-efficacy is related to behavioral intention to use and perceived ease of use of the e-class platform. However, there are is no statistically significant relation between self-efficacy and attitudes towards e-class and perceived usefulness.

Perceived usefulness has statistically significant effects on both attitude towards e-class and behavioral intention to use. More specifically, as is the case in the TAM theory, perceived usefulness was proved to affect behavioral intention affair which was not verified in the study conducted by Park (2009). This could be attributed to the fact that in Korea, students are already familiar with using the Internet in their daily life and, as a consequence, their familiarization with it is a great facilitator in their academic life, while in Greece the use of educational technologies and specifically the use of LMS constitutes a novel practice for students who are not fully accustomed to using them while learning a specific subject. Moreover, this study proves that perceived usefulness influences attitude towards e-class, an effect which is supported by Park's study, because it is considered to be particularly important as the field of marketing. At this juncture, in a positive atmosphere, students themselves create a positive attitude towards this platform that can contribute positively to students re-employing and making avail of a similar platform.

As far as social norm is concerned, it appears to have statistically significant effect on behavioral intention to use, attitude, perceived ease of use and perceived usefulness. Social influence can affect the way users accept a technology and shape their behavior towards it (Gradon, Alshare, & Kwan, 2005). This ascertainment was also reached in part by Park (2009) who confirmed statistically significant effects between social norm and attitude, behavioral intention to use and perceived usefulness. As mentioned previously, social factors affect significantly students in Korea, because in this country, everybody is encouraged to use educational technologies in education.

In Greece, confirmation of these relationships may be explained by the model of modern society, in which technologies play a significant role. Besides this, the fear of exclusion from social environment in case where someone is not technology literate is a factor which plays an important part in young people's behaviors. Subsequently, young people encourage each other to use educational technologies either because they influence each other, or because they do not want to be regarded as 'digitally illiterate', or because they believe that it is this approach will help themselves in their future career path.

Another important factor which was observed to have significant mediation effect is system access, on the ground that it verified the effects among behavioral intention, attitude, perceived ease of use and perceived usefulness. However, the findings for this factor differ with those reported by Park (2009) since the only relationship confirmed was between perceived ease of use. In Korea, a modern web infrastructure has been already developed, thus being very common to students in universities. In consequence, they do not worry about system access, since they know that there are the appropriate facilities and there are not any problems with access. However, in Greece, technology dissemination is less developed than in Korea and this is ascribed to the fact that system access plays an important role in technology acceptance as well as in shaping perceptions about the technology.

Moreover, the effect of attitude towards behavioral intention effect was confirmed, further verifying Park's ascertainment, as the attitude of students to educational technology can shape behavior towards it and eventually lead them to accept it or not. Lastly, the factor of student's academic year, which was introduced for examination in this study, was found to be affected only by perceived ease of use at a significant level. This can be interpretable by the degree to which a person believes that the e-class will be easy το use is influenced by their age, and their accumulated experience with the LMS technology, a finding which is in line with the findings reported by Orfanou, Katsanos and Tselios (2015).

Moreover, it has been shown that perceived ease of use had had statistically significant effect on perceived ease of use and attitudes towards e-class. This effect was also found by Park, as it can be explained through TAM theory, as based on how easy the students believe the system use is, students can accordingly shape an attitude to it altogether. To summarize, the findings were:

*RQ1:* According to the analysis, behavioral intention to use is greatly affected by social influence, system access but also perceived usefulness, self-efficacy and perceived ease of use. However, it was not supported by the hypothesis year and behavioral intention.

*RQ2:* The factor of attitude towards e-class was found to be mainly influenced by system access and then social influence and perceived usefulness but seemed not to be influenced by year and perceived ease of use.

*RQ3:* University students' perceived usefulness of e-class is affected mainly by social norm and system accessibility. However, the effects of perceived ease of use, e- self-efficacy and year were not supported.

*RQ4*: Finally, the hypotheses about students' perceived ease of use were supported which were affected by their e-learning self-efficacy, social norm, system accessibility and academic year.

Moreover, the conclusions underline the usefulness of this research in the educational process at a practical level. The reported findings could help those who create or manage specific learning systems, since they provide important information as to whether students accept or not such systems and their intention for future use. For this reason, teachers and those engaged in the field of LMS should pay attention to factors related to the explanation of behavior and intention of such systems. In addition to the research findings, a usability evaluation was implemented that nowadays is needed in this type of systems, because they can contribute in a simple and quick way to their evaluation and to redevelopment to improve them. More explicitly, the use of such tools could be performed in short period of time, comparing usability of different systems or assessing the various systems which can serve as feedback to optimize the systems so as to be more appealing to the public targeted.

This study is not without limitations. The sample consisted of a single student population with certain characteristics. Thus, participants from other departments and universities located in Greece could participate in order to collect more representative data. Other learning management systems should be also investigated, thus enabling a systematic comparison between them. Finally, these factors could be considered in relation to some characteristics of participants, e.g. using the Big Five Personality Test (Rothmann & Coetzer, 2003). However, a deeper understanding of all learners' cognitive strategies and information processing behaviors is required to provide a suitable information architecture that promotes the learning process (Altanopoulou, Tselios, Katsanos, Georgoutsou, & Panagiotaki, 2015, Katsanos, Tselios, & Avouris, 2008, Tselios, Avouris, & Kordaki, 2002, Tselios, & Avouris, 2003).